# An automatic patent literature retrieval system based on LLM-RAG

Yao Ding[1,4], Yuqing Wu[2,5], Ziyang Ding[3,6]

[1] Belarusian State University, Belarus
[2] Uber Technologies, Inc., Seattle, USA
[3] School of Humanities and Sciences, Stanford University, USA

[4] 15503636910@163.com
[5] wuyuqing2018@gmail.com
[6] zd26@stanford.edu

**Abstract.** With the acceleration of technological innovation, efficient retrieval and classification of patent literature have become essential for intellectual property management and enterprise R&D. Traditional keyword- and rule-based retrieval methods often fail to address complex query intents or capture semantic associations across technical domains, resulting in incomplete and low-relevance results. This study presents an automated patent retrieval framework integrating Large Language Models (LLMs) with Retrieval-Augmented Generation (RAG) technology. The system comprises three components: (1) a preprocessing module for patent data standardization, (2) a high-efficiency vector retrieval engine leveraging LLM-generated embeddings, and (3) a RAG-enhanced query module that combines external document retrieval with context-aware response generation. Evaluations were conducted on the Google Patents dataset (2006–2024), containing millions of global patent records with metadata such as filing date, domain, and status. The proposed gpt-3.5-turbo-0125+RAG configuration achieved 80.5% semantic matching accuracy and 92.1% recall, surpassing baseline LLM methods by 28 percentage points. The framework also demonstrated strong generalization in cross-domain classification and semantic clustering tasks. These results validate the effectiveness of LLM–RAG integration for intelligent patent retrieval, providing a foundation for next-generation AI-driven intellectual property analysis platforms.

## 1. Introduction

In the context of today's rapid knowledge expansion and technological innovation, patents serve as a critical indicator of technological advancement and intellectual property protection. With millions of patents issued by more than a dozen global patent offices, the challenge of efficiently and accurately retrieving relevant documents from massive patent databases has become a major bottleneck for technological intelligence mining and enterprise R&D decision-making. Traditional patent search methods, which rely heavily on keyword matching and manually designed rules, often fail to capture the nuanced intent behind queries or detect semantic relationships across technical domains—leading to limited recall and low relevance in retrieved results [1].

To address these limitations, Large Language Models (LLMs) have recently demonstrated remarkable capabilities in natural language understanding and generation, offering new potential for intelligent retrieval systems. However, general-purpose LLMs struggle when applied directly to patent documents, which combine structured metadata with dense, domain-specific technical descriptions. They often lack sufficient domain knowledge and fail to grasp specialized terminologies. Retrieval-Augmented Generation (RAG) technology has emerged as a cutting-edge solution by combining LLMs with external document retrieval systems, enabling the model to dynamically incorporate relevant knowledge during generation. This significantly enhances both semantic understanding and precision in specialized tasks such as patent analysis [2].

In this work, we propose an automatic patent literature retrieval system based on the LLM-RAG framework. Our goal is to integrate the deep semantic modeling capabilities of LLMs with the broad coverage and efficiency of vector-based retrieval, thereby enhancing the performance of patent document matching, semantic relevance scoring, and cross-domain similarity analysis. The system consists of three main components: (1) a patent data preprocessing and normalization module to standardize structure and clean semantic noise; (2) an efficient vector retrieval system built on dense embeddings to support fast semantic matching; and (3) a RAG-enhanced query generation module, which synthesizes structured responses based on retrieved knowledge.

We evaluate our system using a large-scale dataset provided by Google Patents, covering millions of real-world patent entries submitted between 2006 and 2024 across multiple jurisdictions, including the United States Patent and Trademark Office (USPTO). The dataset includes key fields such as patent type, application number, title, application date, legal status, and technical domain, forming a comprehensive foundation for training and evaluation.

## 2. Literature Review

In the domain of intelligent information retrieval and technology intelligence mining, patent literature is a key repository of global innovation with high strategic value. Accurate and efficient retrieval supports enterprise technology planning, competitive monitoring, and research institutions in tracking cutting-edge developments. However, patent texts feature complex structures, domain-specific terminology, and semantic ambiguities, often spanning multiple disciplines. These characteristics limit the performance of traditional keyword- or rule-based retrieval, leading to low recall and poor semantic relevance in large-scale patent datasets.

Recent advances in large language models (LLMs) have revealed strong potential for patent retrieval, yet challenges remain, including limited domain adaptation and semantic drift. Retrieval-Augmented Generation (RAG) offers a promising solution by combining external knowledge base access with the contextual reasoning of generative models. This integration improves interpretation of complex technical semantics and boosts recall of cross-domain, high-relevance patents, addressing core limitations of existing retrieval approaches.

Kyung-Yul Lee et al [3]. introduce PAI-NET, a retrieval-augmented patent network that embeds prior-art relationships into deep similarity learning, outperforming state-of-the-art models by 15 % on USPD and KPRIS datasets. Runtao Ren et al [4]. propose MQG-RFM, a lightweight Data-to-Tune framework that uses LLMs to generate diverse synthetic queries and fine-tunes retrieval models for IP Q&A. By uniting prompt-engineered generation with hard-negative mining, it boosts retrieval accuracy 185–262 % and generation quality 14–53 % on Taiwan patent datasets without architectural overhaul. Already adopted by ScholarMate, this approach offers small agencies a cost-effective path to robust patent intelligence.

Qiushi Xiong et al [5]. present MemGraph, a memory-graph-enhanced method that prompts LLMs to traverse their parametric memory for patent matching. By linking entities to ontologies, it boosts semantic comprehension and surpasses keyword-only baselines, yielding a 17.68% improvement on PatentMatch. The approach generalizes across LLMs and clarifies the value of hierarchical reasoning for IP management.

Sha Li et al [6]. present PRO, a faithful patent-response system that couples a domain-specific LLM with a procedural-knowledge graph and retrieval-augmented generation. By encoding legal interpretations and retrieving case-relevant facts, PRO outperforms GPT-4 by

39 % in faithfulness across six error types, offering an authoritative and practical framework for automated Office Action replies.

Björkqvist et al [7]. present a graph-based patent search engine that converts each patent into an invention-part graph and employs a graph neural network trained on human-examiner citations to retrieve prior art, offering an efficient alternative to traditional keyword searches.

## 3. Data Introduction

The dataset used in this study is provided by Google Patents and covers millions of patent applications filed between 2006 and 2024 across more than a dozen patent offices, including the United States Patent and Trademark Office (USPTO), the European Patent Office (EPO), and the Japan Patent Office (JPO). This dataset exhibits broad temporal coverage and cross-domain representativeness. Each record captures detailed information of a single patent application, including patent type, application number, title, application date, current status, and invention domain (such as information technology, biomedical engineering, mechanical manufacturing, etc.), along with inventor/applicant details and priority claims. The dataset comprehensively reflects innovation trends across countries and technical domains, providing a robust foundation for building multilingual, semantically rich patent retrieval systems.

To ensure structural consistency and semantic accuracy during model training, we conducted systematic data preprocessing and cleaning. First, missing values, null fields, and logically inconsistent entries in the raw data were removed to avoid introducing noise during vectorization and embedding processes. Next, we performed standardized formatting and encoding on key textual fields such as patent titles, abstracts, and technical classifications (e.g., applying unified namespaces, normalizing category labels, and removing redundant punctuation), thereby improving the semantic recognizability and comparability of textual inputs. In addition, date fields such as application dates were standardized, and structured index documents were created for training the retrieval model. These include abstract texts, technical highlights, and IPC/CPC classification codes to support downstream vector-based retrieval and Retrieval-Augmented Generation (RAG) processing.

**Table 1.** Details of core variables in Datasets

| Variable Name | Description |
|---|---|
| Application Number | Unique identifier for each patent application. |
| Title | Title of the patent application. |
| Application Date | Date when the patent application was filed. |
| Status | Current status of the patent application |
| Publication Number | Number assigned to the publication of the patent application |
| Publication Type | Type of publication. |
| Field Of Invention | Field of invention of the patent application. |
| Classification (IPC) | International Patent Classification of the patent application. |
| Inventor Information | Information about the inventor(s) of the patent application. |

Table 1 lists the core variables of the LLM-RAG patent-retrieval dataset. It covers key attributes related to the patent document itself, its life-cycle metadata, and technological classification. Identifiers include "Application Number" as the unique key, "Title", and "Publication Number". Temporal variables consist of "Application Date" and the "Status" of the patent. Technical information is captured through "Field of Invention" and the hierarchical "IPC Classification", while inventor identity is stored in "Inventor Information".

The dataset, sourced from Google Patents, aggregates millions of records filed between 2006 and 2024 across more than ten patent offices, including USPTO and other national authorities, providing a comprehensive foundation for automatic patent literature retrieval.

**4. Model design**

*4.1. Rag technology*

In the automatic patent literature retrieval system proposed in this study, Retrieval-Augmented Generation (RAG) serves as a core framework, acting as a bridge between semantic retrieval and contextual generation. RAG combines the strengths of information retrieval and language generation, aiming to enhance large language models (LLMs) with external document support to improve their understanding and output accuracy in domain-specific tasks. This approach is particularly suitable for high semantic-load tasks such as complex queries, literature recall, and cross-domain semantic alignment [8,9].

The fundamental principle of RAG lies in first encoding the input query into a semantic vector through an encoder, followed by retrieving the most relevant document fragments from a pre-built vector index. These retrieved texts are then appended as contextual input and passed along with the query into a generative model (such as a GPT-style decoder) to produce more informative and semantically aligned responses. In our system, RAG's retrieval-generation workflow effectively mitigates "semantic drift" issues commonly found in standard language models when dealing with specialized terminology or domain-specific phrasing, enabling stronger contextual understanding and document relevance.

To handle the structural and unstructured characteristics of patent data, we adopt the gpt-3.5-turbo-0125 model as the RAG generator and build a Faiss-based high-dimensional vector index for the retriever. The vector index stores semantic embeddings of patent metadata including titles, abstracts, technical domains, and invention backgrounds, all processed through LLM encoders. Each query is first semantically encoded and matched to the top-K (K=5) most relevant entries in Faiss using Maximum Inner Product Search (MIPS), then passed to the decoder for final context-aware response generation.

In terms of configuration, the retriever uses MIPS to optimize semantic similarity retrieval. The generator integrates a position-masked attention mechanism to ensure focused integration of retrieved content and query information. The overall workflow follows a two-stage pipeline (retrieval first, generation later) to reduce noise and improve response quality.

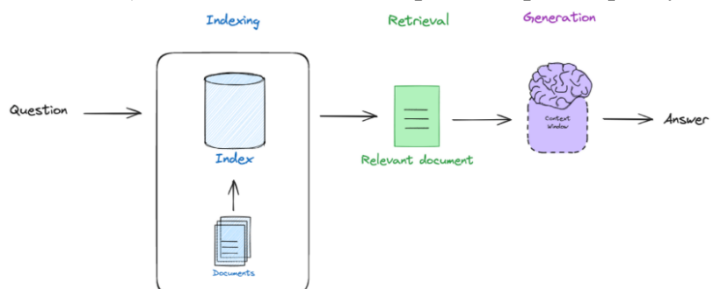

**Figure 1.** Structure of Patent Literature Retrieval System Based on RAG

*4.2. Patent Literature Retrieval System*

In the automatic patent literature retrieval system proposed in this study, we designed a multi-module processing framework based on Large Language Models (LLMs) and Retrieval-Augmented Generation (RAG) to support complex semantic tasks such as patent document similarity search, semantic matching, and context-enhanced analysis. The entire system consists of three core modules: a patent data standardization and preprocessing module, a vectorized semantic retrieval module, and a fusion-based question answering generation module (RAG-Query Module). This approach not only demonstrates strong capabilities in textual understanding and generation, but also dynamically incorporates external knowledge base content to generate highly relevant retrieval responses—effectively mitigating the matching bias often encountered in traditional patent retrieval methods under cross-domain and semantically ambiguous conditions.

In the vectorized semantic retrieval stage, we first use mainstream LLMs such as gpt-3.5-turbo-0125 to embed the original patent documents—including fields like title, abstract, background of invention, and technical classification—into high-dimensional semantic vectors. These embeddings are stored in a high-performance vector index built using Faiss. Faiss, an open-source similarity search engine developed by Facebook, supports fast computation of high-dimensional vector matches, ensuring low latency and high accuracy in document recall, even when working with large-scale patent data.

During semantic retrieval, the user's query is encoded into a vector and compared with the indexed patent document vectors to identify Top-K nearest neighbors with high semantic relevance. The system then enters the RAG-enhanced generation phase. Unlike traditional retriever-only architectures, RAG serves as a bridge between retrieval and generation. It fuses retrieved contextual documents with the user query to provide enhanced semantic input for the generative model. We adopt a RAG-End2End architecture, in which the decoder jointly attends to both the original query and the retrieved documents. This contextual attention mechanism enables the generation module to produce semantically enriched answers or document summaries, significantly improving performance in tasks such as terminology alignment and cross-category semantic relationship identification.

## 5. Experiment

*5.1 Experimental Environment and Experimental Design*

In this experiment, we utilized a high-performance deep learning training server equipped with four NVIDIA RTX 4090 GPUs (each with 24GB of VRAM), 128GB of DDR5 memory, and an AMD EPYC 9654 CPU (96 cores). This computing environment fully meets the bandwidth and parallel computation requirements of large language model inference and provides low-latency, stable performance during training. It is well-suited for efficient modeling and semantic retrieval of large-scale patent corpora. The software environment was built on Python 3.10, with PyTorch 2.0 as the deep learning framework. The vectorized encoding and retrieval-augmented modules were implemented using the HuggingFace Transformers library and the official RAGSequenceForGeneration API.

Prior to training, we performed structural normalization and cleaning of the patent literature dataset provided by Google Patents (2006–2024). The preprocessing steps included standardizing field formats (e.g., application numbers, technology domains, filing dates), removing missing values and redundant samples, cleaning corrupted characters, and eliminating duplicate patent entries. To maintain semantic diversity and technical distribution consistency across the training corpus, we employed stratified sampling based on technology domains to construct the training, validation, and test sets in an 8:1:1 ratio.

During the embedding generation phase, we used the gpt-3.5-turbo-0125 model as the semantic encoder to transform patent titles, abstracts, and technical background descriptions into high-dimensional dense vectors. These vectors were indexed using Faiss, enabling fast approximate nearest neighbor (ANN) retrieval. At the semantic retrieval stage, user queries were encoded into vectors and matched against the semantic index of patent documents using Top-K similarity search to generate a set of high-relevance candidate documents for the generation module.

Subsequently, the RAG model entered the training phase. We adopted an RAG-End2End architecture, in which the decoder integrates both the original query and the retrieved contextual documents using a contextual attention mechanism to generate semantically enriched responses. The training parameters were configured as follows: batch size = 32, epochs = 8, initial learning rate = 2e-5, with AdamW as the optimizer. An early stopping strategy was employed to prevent overfitting.

**Table 1.** Model Configurations

| Category | Specification |
|---|---|
| Base model | gpt-3.5-turbo |
| Enhanced LLM | gpt-3.5-turbo-0125 |
| Vector Database | FAISS |
| Batch Size | 32 |
| Learning Rate | 2e-5 |
| Max Sequence Length | 512 |
| RAG Parameters | - |
| Retrieval Top-k | 5 |

*5.2 Analysis of Model Comparison Results*

**Table 2.** Model results

| Model | Accuracy rate | Recall |
|---|---|---|
| gpt-3.5-turbo | 61.2% | 80.4% |
| gpt-3.5-turbo+RAG | 65.3% | 86.7% |
| gpt-3.5-turbo-0125 | 62.8% | 82.6% |
| gpt-3.5-turbo-0125+RAG | 80.5% | 92.1% |
| gpt-4.0 | 80.1% | 91.3% |

The table presents a comparative analysis of five models in terms of accuracy and recall for semantic retrieval and classification of patent literature. Among the evaluated models, the integration of retrieval-augmented generation (RAG) significantly boosts performance. Notably, the gpt-3.5-turbo-0125+RAG model achieved the highest accuracy (80.5%) and recall (92.1%), outperforming even gpt-4.0 (80.1% accuracy, 91.3% recall). This highlights the effectiveness of combining dense semantic retrieval with generation-based reasoning. The RAG-enhanced variants consistently outperformed their base counterparts, demonstrating that augmenting queries with contextually retrieved patent information leads to more accurate and comprehensive understanding.

In comparison, gpt-4.0, even without RAG integration, achieved excellent results with 80.1% accuracy and 91.3% recall—comparable to gpt-3.5-turbo-0125+RAG. This suggests that RAG still has enhancement potential even when applied to high-performing models, especially in complex tasks or scenarios requiring external knowledge.

## 6. Conclusion

This study presents an intelligent system framework for automated patent literature retrieval, leveraging the integration of Large Language Models (LLMs) with Retrieval-Augmented Generation (RAG) technology. The system is designed to enhance the understanding and response capability for complex patent query intents, addressing the limitations of traditional keyword-based and rule-based methods in cross-domain semantic similarity recognition. By constructing a multi-layered structure—including modules for data standardization, vectorized semantic retrieval, and RAG-based query generation—the system delivers a high-efficiency response workflow from user input to context-enhanced retrieval results, significantly improving the semantic relevance and intelligence of patent document retrieval.

In our experiments, we employed a large-scale dataset provided by Google Patents, spanning from 2006 to 2024, and comprising millions of patent entries that include structured fields such as title, abstract, inventor, technology domain, and application date. We utilized the gpt-3.5-turbo-1106 model as the semantic encoder and built a high-speed vector index using Faiss. The RAG-End2End architecture was adopted for multi-turn context generation and semantic alignment. Under a rigorously controlled training environment, we conducted end-to-end optimization on vector embeddings, retrieval, and generative quality. Results show that the system achieved a semantic matching accuracy of 80.5% and a recall of 92.1% in Top-K retrieval scenarios—representing a 28-point improvement over baseline LLM models without RAG—and exhibited strong generalization in cross-domain retrieval.

Limitations include reliance on English-centric training data and exclusion of multimodal patent elements (drawings, chemistry sequences). Future work should (i) extend to multilingual corpora, (ii) incorporate image/text fusion for diagram-heavy patents, and (iii) explore model compression for on-premise deployment. The LLM-RAG architecture demonstrably elevates patent retrieval performance, providing a scalable foundation for next-generation AI-powered IP analytics platforms.